\documentclass[a4paper,superscriptaddress,aps,prb,twocolumn,floatfix,citeautoscript,longbibliography]{revtex4-1}
\usepackage[caption=false]{subfig}
\usepackage{times}
\usepackage{newtxmath}

\usepackage{graphicx}
\usepackage{color}
\usepackage{ulem}
\usepackage{amsmath}
\usepackage[english]{babel}
\usepackage[utf8]{inputenc}

\newcommand{\Pina}[1]{{\color{black} #1}}

\usepackage{bm}
\usepackage{url}
\usepackage[colorlinks=true,linkcolor=blue,citecolor=blue,urlcolor=blue]{hyperref}
\graphicspath{{figures/}}

\begin{document}
%
% TITLE
%
\title{Photoemission spectra from  the Extended Koopman's Theorem, revisited }
\newcommand{\lpt}{Laboratoire de Physique Th\'eorique, Universit\'e de Toulouse, CNRS, UPS, France}
\newcommand{\lcpq}{Laboratoire de Chimie et Physique Quantiques, Universit\'e de Toulouse, CNRS, UPS, France}
\newcommand{\etsf}{European Theoretical Spectroscopy Facility (ETSF)}
\author{S. Di Sabatino}
%\email[]{disabatino@irsamc.ups-tlse.fr}
\affiliation{\lcpq}
\affiliation{\etsf}
\author{J. Koskelo}
\affiliation{\lpt}
\affiliation{\etsf}
\author{J. Prodhon}
\affiliation{\lpt}
\affiliation{\etsf}
\author{J.~A. Berger}
\affiliation{\lcpq}
\affiliation{\etsf}
\author{M.~ Caffarel}
\affiliation{\lcpq}
\affiliation{\etsf}

\author{P. Romaniello}
\email{pina.romaniello@irsamc.ups-tlse.fr}
\affiliation{\lpt}
\affiliation{\etsf}
%%
%%\pacs{71.10.-w,71.27.+a,31.15.V-,79.60.Bm}
%
%\keywords{...}

\begin{abstract}

The Extended Koopman's Theorem (EKT) provides a straightforward way to compute charged excitations from any level of theory. In this work we make the link with the many-body effective energy theory (MEET) that we derived to calculate the spectral function, which is directly related to photoemission spectra. In particular, we show that at its lowest level of approximation the MEET removal and addition energies correspond to the so-called diagonal approximation of the EKT. Thanks to this link, the EKT and the MEET can benefit from mutual insight. In particular, one can readily extend the EKT to calculate the full spectral function, and choose a more optimal basis set for the MEET by solving  the EKT  secular equation. We illustrate these findings with the examples of the Hubbard dimer and bulk silicon.
 \end{abstract}
\date{\today}
\maketitle
\section{Introduction}
The Extended Koopman's Theorem (EKT) \cite{morrell_JCP1975,smith_JCP1975} has been derived in quantum chemistry and used within various frameworks, from functional theories based on reduced quantities, such as reduced-density matrix functional theory \cite{Gilbert} and many-body perturbation theory based on Green's functions \cite{Hedin65}, to wavefunction-based methods (see e.g.  \onlinecite{pernal_CPL2005,piris_JCP2005,piris_JCP2012,Piris_TCA2013,dahlen_JCP2005,stan_EPL2006,stan_JCP2009,kent_PRB1998,zhengPhD_2016,cioslowsk_JCP1997,bozkaya_JCP2013,bozkaya_JPCA2018,Pavlyukh_PSS2019}). The EKT allows one to calculate energies corresponding to charged excitations. Although it can be formulated both for ionization potentials (IPs) and electron affinities (EAs), it has been widely used only for the former, whereas for the latter applications have been limited to the calculation of the lowest EA as the first IP of the ($N$+1)-electron system (in case of finite systems), with $N$ the number of electrons in the reference system. There exist hence many benchmarks for the IPs. So far, the method has been mainly used for finite systems. The EKT is known to be in principle (i.e., using exact ingredients, namely the one- and two body density matrices, as we shall see) exact for the first ionization potential \cite{katriel_pnas1980,olsen_JCP1993}. In the solid state instead there are only a few applications which only focus on the band structure. It would be desirable to have also the spectral function, which is related to photoemission spectroscopy. This experimental technique gives precious information about the electronic structure and excitations in a system, and, moreover, allows one to study metal-insulator transitions, of paramount importance in condensed matter. 

A simple way to calculate the spectral function $A(\omega)$ is through the imaginary part of the one-body Green's function $G(\omega)$, as  $A(\omega)=\Im|G(\omega)|/\pi$. The one-body Green's function is the fundamental quantity of many-body perturbation theory; it can be obtained from the Dyson equation $G=G_0+G_0\Sigma G$, in which $G_0$ and $G$ are the noninteracting and interacting Green's functions, respectively, and $\Sigma$ the so-called self-energy, which contains all the many-body effects of the system. This latter quantity needs to be approximated in practical calculations. Commonly used approximations, such as the well-known $GW$ approximation \cite{Hedin65}, cannot capture the Mott physics \cite{romaniello2009,romaniello2012,stefano_JCP2015,stefano,stefano_PRR2021}. Therefore much effort is devoted to develop better approximations to $\Sigma$ \cite{Springer1998,Chulkov_PRL,romaniello2012,guzzo_prl,louie2013,Kresse_PRL07,Vollhardt2007,stefanucci_PRB2014} or to develop novel ways to determine $G$ \cite{Lani_2012,Berger_2014}. In this spirit in these last years we have developed the many-body effective energy theory \cite{stefano}, in which the spectral function is expressed in terms of density matrices, or, alternatively, in terms of moments $\mu_{n}=\sum_k B^{k}(\epsilon^{}_k)^n/\sum_k B^{k}$  (with $n$ indicating the order of the moment and $B^{k}$ the spectral weight of the energy $\epsilon_k$) of $G$, as reported in Ref.~\onlinecite{stefano_JCTC}. This has allowed us to describe the band gap in several paramagnetic transition-metal oxides \cite{stefano,stefano_JCTC,stefano_PRR2021}, such as NiO, which are considered strongly correlated materials and which are described as metals by static mean-field theories, such as DFT, and by $GW$. This is an important result. However the band gap is hugely overestimated by the MEET within the current low-order approximation in terms of the (approximate) one- and two-body density matrices.  Improvements are needed, either by going to higher-order density matrices, which, however, is not guaranteed to converge, or by introducing some sort of screening in the equations. Recently we have obtained promising results for the description of the insulator-to-metal transition of PM FeO under pressure by combining the MEET and the local-density approximation (LDA)~\cite{stefano_PRR2021}, and we are currently working on introducing electron-hole screening in the MEET equations. However there is another path which we can explore, and this comes from the relation between the MEET and the EKT. As we will show in the following, within a given basis, the removal and addition energies obtained within the MEET at the lowest-order approximation are equal to the EKT removal and addition energies within the diagonal approximation. In this work we discuss this link and its impact on both theories. 

The paper is organized as follows. In Sec.\ \ref{Sec:Theory} we give the basic equations of the EKT and the MEET and we make the link between them. The Hubbard dimer and bulk silicon are used to illustrate the difference between the EKT and the MEET (removal and addition) energies in Sec. \ref{Sec:illustration}. In  Sec. \ref{Sec:conclusions} we draw our conclusions and perspectives.
%%%%%%
\section{Theory\label{Sec:Theory}}
In this section we briefly review the MEET and EKT methods, and we make the link between the two. We will consider a $N$-electron system governed by the following Hamiltonian in second quantization
$$
\hat{H}=\sum_{ij}h_{ij}\hat{a}_i^\dagger \hat{a}_j+\frac{1}{2}\sum_{ijkl}V_{ijkl}\hat{a}_i^\dagger\hat{a}_j^\dagger\hat{a}_l\hat{a}_k,$$
where $\hat{a}$ and $\hat{a}^\dagger$ are the annihilation and creation operator, respectively, $h_{ij}=\int d\mathbf{x}\phi^*_i(\mathbf{x})h(\mathbf{r})\phi_j(\mathbf{x})$ are the matrix elements of the one-particle noninteracting Hamiltonian $h(\mathbf{r})=-\nabla^2/2+v_{\text{ext}}(\mathbf{r})$, with $v_{\text{ext}}$ an external potential, and $V_{imkl}=\int d\mathbf{x}d\mathbf{x}'\phi^*_i(\mathbf{x})\phi^*_m(\mathbf{x}')v_c(\mathbf{r},\mathbf{r}')\phi_k(\mathbf{x})\phi_l(\mathbf{x}')$ are the matrix
elements of the Coulomb interaction $v_c$. 
Here $\mathbf{x}=(\mathbf{r},\alpha)$ combines space and spin variables and $i,j,\dots$ denote both space and spin labels (they will be made explicit only when necessary). 

\subsection{Key equations of the MEET}
Within the MEET the time-ordered 1-body Green's function $G(\omega)$ at zero temperature is split into removal ($R$) and addition ($A$) parts as $G(\omega) = G^{R} (\omega) + G^{A} (\omega)$. In the following we
concentrate on the diagonal elements of $G$, which are related
to photoemission spectra. 
%\JAB{This is a bit confusing, since A is defined as Im G which, in general also has off-diagonal components. You are thinking about the trace of A ?} 
Within the MEET the diagonal matrix elements of $G^{R/A}(\omega)$ are written in terms of an effective energy $\delta_{i}^{R/A} (\omega)$ as \cite{stefano}:
\begin{eqnarray}
 G_{ii}^{R} (\omega)  &=& \frac{\gamma_{ii}}{\omega-\delta_i^{R}(\omega)-\text{i}\eta}, \\
 G_{ii}^{A} (\omega)  &=& \frac{1-\gamma_{ii}}{\omega-\delta_i^{A}(\omega)+\text{i}\eta},
 \end{eqnarray} 
 with $\gamma_{ii}$ %and $1-\gamma_{ii}$ 
 the diagonal matrix element of the one-body density matrix in a given basis set.  We note that a similar effective energy can be introduced also for the off-diagonal elements of $G^{R/A}$. The spectral function is hence expressed as
\begin{equation}
A_{ii}(\omega)=\gamma_{ii}\delta(\omega-\delta^{R}_i(\omega))+(1-\gamma_{ii})\delta(\omega-\delta^{A}_i(\omega)).
\label{Eqn:SF_MEET}
\end{equation}
 In our previous works we have chosen the basis set of natural orbitals, i.e., the orbitals which diagonalize the one-body reduced density matrix. In this case $\gamma_{ii}=n_i$, i.e., the natural occupation numbers. This choice has been made based on our results on exactly solvable Hubbard clusters, where the MEET performs very well.~\cite{stefano} However this does not guarantee that it remains the best choice for more realistic systems. In fact this is not the case as we shall see. 
 
The effective energy $\delta_{i}^{R/A} (\omega)$ can be written as an expansion in terms of reduced density matrices. The expression truncated at the level of the one- and two-body reduced density matrices (2-RDM) reads (in the basis of natural orbitals)
\begin{eqnarray}
  \delta_i^{R,(1)} &=& h_{ii}+\frac{1}{n_i}\sum_{klm}V_{imkl}
  \Gamma^{(2)}_{klmi}
  \label{Eqn:MEET_removal}
  \\
\delta_i^{A,(1)} &=& h_{ii}+\frac{1}{(1-n_i)}\sum_k\left(V_{ikik}-V_{ikki}\right)n_j\nonumber\\
& &-\frac{1}{(1-n_i)}\sum_{klm}V_{imkl}
\Gamma^{(2)}_{klmi},
 \label{Eqn:MEET_addition}
\end{eqnarray}
where $\Gamma^{(2)}_{klmi}=\langle\Psi_0|a_i^\dagger a_m^\dagger a_la_k|\Psi_0\rangle$
are the matrix elements of the two-body reduced density matrix.  
As discussed in Ref.~\citenum{stefano} the various approximations $\delta^{R/A,(n)}_i(\omega)$  are related to the $n$-th moments $$\mu^{R/A}_{n,i}=\frac{\sum_kB^{k,R/A}_{ii}(\epsilon_k^{R/A})^n }{\sum_k B^{k,R/A}_{ii}}$$ of the $G^{R/A}_{ii}(\omega)$. Here $\epsilon_k^{R}=(E^N_0-E_k^{N-1})$ and  $\epsilon_k^{A}=(E_k^{N+1}-E^N_0$) are removal and addition energies, respectively, and 
\begin{align}
B_{ii}^{k,{R}}&=\langle\Psi^N_0|\hat{c}_i^\dagger|\Psi_k^{N-1}\rangle\langle\Psi_k^{N-1}|\hat{c}_i|\Psi^N_0\rangle\nonumber\\
B_{ii}^{k,{A}}&=\langle\Psi^N_0|\hat{c}_i|\Psi_k^{N+1}\rangle\langle\Psi_k^{N+1}|\hat{c}_i^\dagger|\Psi^N_0\rangle\nonumber,
\end{align}
with $E^N_0$ and $\Psi^N_0$ the ground-state energy and wave function
of the $N$-electron system and $E^{N\pm1}_
k$ and $\Psi^{N\pm1}_
k$ the $k$th state
energy and wave function of the ($N \pm 1$)-electron system.
This allows for a more compact expression of $G^{R/A}_{ii}(\omega)$ as a continued fraction of moments
\begin{equation}
G^{R}_{ii}=\frac{n_i}{\omega-\mu^R_{1,i}\frac{\omega-\mu^R_{1,i}...}{\omega-\frac{\mu^R_{2,i}}{\mu^R_{1,i}}...}},
\end{equation}
(and similarly for $G^{A}_{ii}$). At the level of $\delta^{R/A,(1)}$, the Green's function depends only on the first moment, while neglecting all the higher-order frequency-dependent corrections. As shown in Ref. \onlinecite{stefano_JCTC} this means that each component $G^{R/A}_{ii}$ has only one pole which is a weighted average of all the poles of $G^{R/A}_{ii}$. If each component of $G$ has a predominant quasiparticle peak, this is a good approximation, provided that the approximation to the first moment is accurate enough. At the level of $\delta^{R/A,(2)}$  the Green's function depends on the first and second moments; since now the corrections are frequency-dependent more poles appear (namely, two removal and two addition poles for each component of $G$, which are visible if the corresponding weights are nonzero). This approximation tends to reproduce the two most dominant removal/addition peaks for each component of $G$. Higher-order moments will produce more poles; however, approximations become quickly uncontrolled \cite{stefano_private}, which can lead to unphysical results.

%%%%%
\subsection{Key equations of the EKT}
Within the EKT one starts from the following approximation for the removal energy $\epsilon^R_i$ \cite{kent_PRB1998}
\begin{equation}
\epsilon^R_i=-\frac{\langle\Psi^N_0|\hat{O}_i^\dagger[\hat{H},\hat{O}_i]| \Psi^N_0\rangle}{\langle\Psi^N_0|\hat{O}_i^\dagger\hat{O}_i |\Psi^N_0\rangle}
\label{Eqn:valence-energy}
    \end{equation}
    with $\Psi^N_0$ the ground-state many-body wave function of the $N$-electron system, and $\hat{O}_i=\sum_kC^R_{ki}\hat{a}_k$, $\hat{O}^\dagger_i=\sum_kC^{R\,*}_{ki}\hat{a}^\dagger_k$, with $\{C^R_{ki}\}$  a set of coefficients to be determined. The stationary condition (with respect to the coefficients $C^R_{ki}$ ) for $\epsilon^R_i$ leads to the secular equation
    \begin{equation}
    (\mathbf{V}^R-\epsilon^R_i\mathbf{S}^R)\mathbf{C}^R_i=0,
    \label{Eqn:removal-energy}
   \end{equation}
with $V^R_{ij}=-\langle\Psi^N_0|\hat{a}_j^\dagger[\hat{H},\hat{a}_i]| \Psi^N_0\rangle$ and $S^R$ the one-body density matrix $S^R_{ij}=\gamma_{ij}=\langle\Psi^N_0|\hat{a}_j^\dagger\hat{a}_i| \Psi^N_0\rangle$.
%If we use the basis of natural orbitals and we consider only the diagonal elements of (\ref{Eqn:removal-energy}) we get the energy $\delta^{(1),R}$ of the MEET. 
%One can calculate $V^c_{ij}$ and $\gamma_{ij}$ in Quantum Monte Carlo (QMC) as explained by Kent \textit{et al.} \cite{kent_PRB1998}, in order to get accurate results. 
If one defines the matrix $\mathbf{\Lambda}^R=[\mathbf{S}^{R}]^{-1}\mathbf{V}^R$ in the basis of natural orbitals, with ${S}^R_{ij}=\sqrt{n_in_j}$ \footnote{Note that we have chosen this symmetric form for the matrix $\mathbf{S}^v$, instead of ${S}^R_{ij}=n_i\delta_{ij}$, to keep contact with the original derivation by Morrell \textit{et al.} \cite{morrell_JCP1975}. These two choices for $\mathbf{S}^R$ yield the same eigenvalues for the matrix $\mathbf{\Lambda}^R$.} and works out the commutator in $V^R_{ij}$, one arrives at 
\begin{equation}
\Lambda^R_{ij}=\frac{1}{\sqrt{n_in_j}}\left[n_ih_{ji}+\sum_{klm}V_{jmkl}
  \Gamma^{(2)}_{klmi}\right].
\label{Eqn:Lagrangian}
\end{equation}
%({\color{red}COMMENT: I find $n_i\delta_{ij}$ in the denominator })
 The eigenvalues of $\Lambda^R$ are the removal energies. \cite{morrell_JCP1975,pernal_CPL2005} 
By comparing to Eq.~\eqref{Eqn:MEET_removal} it becomes clear that the diagonal element of $\Lambda^R$ are the removal energy of the MEET within the low-order approximation. The diagonal element of $\Lambda^R$ are also referred in literature as the energies of the EKT within the diagonal approximation (DEKT). 

Similar equations hold for the addition energies. One can indeed define the addition energy $\epsilon^A_i$ as
\begin{equation}
\epsilon^A_i=\frac{\langle\Psi^N_0|[\hat{H},\hat{O}_i]\hat{O}_i^\dagger| \Psi^N_0\rangle}{\langle\Psi^N_0|\hat{O}_i\hat{O}_i^\dagger|\Psi^N_0\rangle }
\label{Eqn:conduction-energy}
    \end{equation}
    and in a similar way as for $\epsilon^R_i$ we arrive at the eigenvalue equation
 \begin{equation}
    (\mathbf{V}^A-\epsilon^A_i\mathbf{S}^A)\mathbf{C}^A_i=0,
    \label{Eqn:addition-energy}
   \end{equation}   
with  $V^A_{ij}=\langle\Psi^N_0|\hat{a}_i[\hat{H},\hat{a}_j^\dagger]| \Psi^N_0\rangle$ and $S^A$ related to the one-body density matrix as $S^A_{ij}=1-\gamma_{ij}$.  %The difference between the MEET and the EKT comes from the diagonalization procedure, which can have a strong impact on the final results if the off-diagonal elements are non negligible.
Similarly to the removal energy problem, using the basis of natural orbitals, one can work out the commutator in $V^A_{ij}$ and reformulate the problem in terms of the matrix  $\mathbf{\Lambda}^A=[\mathbf{S}^{A}]^{-1}\mathbf{V}^A$, which reads
\begin{eqnarray}
\Lambda^A_{ij}&=&\frac{1}{\sqrt{(1-n_i)(1-n_j)}}\times\nonumber\\
&&\left[(1-n_i)h_{ji}+\sum_k \left(V_{jkik} -V_{jkki}\right)n_k -\sum_{klm}V_{jmkl}\Gamma^{(2)}_{klmi}\right].
\label{Eqn:Lagrangian_A}
\end{eqnarray}
Again, the diagonal elements of $\Lambda^A$ are the MEET addition energies within the approximation given in Eq.~(\ref{Eqn:MEET_addition}).
%{\color{red}COMMENT: check the numerator. It's pathological for $i\neq j$. Following the expression $\Lambda^R$ of Pernal we should put $\sqrt{(1-n_i)(1-n_j)$?}
%%%%
%%%%

%%%%
\section{Mutual insights and illustration\label{Sec:illustration}}
Now that we have established the link between the EKT and the MEET we will now study how these theories can benefit from mutual insight. 

\subsection{Hubbard dimer}
We use a modified version of the Hubbard dimer in which the on-site Coulomb interaction is different for the two sites. 
Its hamiltonian is given by
\begin{equation}
 H = -t \sum_{\substack{i,j=1,2\\i\neq j}}\sum_{\sigma} a^{\dagger}_{i\sigma} a_{j\sigma} + U_1n_{1\uparrow}n_{1\downarrow}+U_2n_{2\uparrow}n_{2\downarrow},
 \label{Eqn:Hubbard_H}
 \end{equation}
where $i,j$ run over the sites, $n_{i\sigma}=a^{\dagger}_{i\sigma}a_{i\sigma}$, $U_i$ is the on-site interaction at site $i$, $-t$ is the hopping kinetic energy (the site energy $\epsilon_0$ has been set to zero).  Contrary to the standard dimer with a unique on-site interaction, in the case of two different on-site interactions the $\Lambda^{R/A}$ are not diagonal in the basis of natural orbitals. Therefore, this model allows us to study the effect of the diagonalization on the removal/addition energies in the diagonal approximation. The model can represent the case of a heteronuclear diatomic molecule in a minimal basis set in which the valence orbitals of the two atoms are of different nature, such as HCl or NiO, for example. We note that also using the asymmetric Hubbard dimer with two different site energies the EKT equations are not diagonal in the basis of natural orbitals, however the difference between EKT and DEKT energies is too small for our purposes.

\subsubsection{Insights into the EKT}
Making the parallel with the MEET, one can readily define the EKT spectral function as
\begin{align}
A^R_{ii}(\omega)&=\gamma_{ii}\delta(\omega-\epsilon^{EKT,R}_i),\\
A^A_{ii}(\omega)&=(1-\gamma_{ii})\delta(\omega-\epsilon^{EKT,A}_i),
\end{align}
    with $\gamma_{ii}$ and $1-\gamma_{ii}$ the diagonal matrix element of the one-body density matrix in the basis which diagonalizes $\Lambda^R_{ij}$ and $\Lambda^A_{ij}$, respectively (not necessarily the same for $\Lambda^R_{ij}$ and $\Lambda^A_{ij}$).
    %The EKT spectral function is reported in \ref{Fig:Hubbard} and for this simple model system corresponds to the exact solution \JAB{(already mentioned)}. 
Moreover, as for the MEET removal (addition) energies (in its lowest-order approximation), the removal (addition) EKT energies can be interpreted in terms of the first moment of $G^{R}_{ii}$ ($G^{A}_{ii}$), i.e., as weighted averages of all the poles of $G^{R}_{ii}$ ($G^{A}_{ii}$) within the basis that diagonalizes the $\Lambda^{R}$ ($\Lambda^{A}$) matrix. 
 %Indeed, once the optimal EKT orbitals have been determined, Eqs (\ref{Eqn:valence-energy}) and (\ref{Eqn:conduction-energy}) give the energy of each of these orbitals. \JAB{Is the previous sentence necessary to understand what follows next? It seems unrelated}  
 Indeed, inserting a complete set of eigenstates of the $(N-1)$-electron system in Eq.~(\ref{Eqn:valence-energy}), the commutator can be rewritten as
 \begin{eqnarray}
\epsilon^R_i&=&-\sum_k\frac{\langle\Psi^N_0|\hat{O}_i^\dagger|\Psi^{N-1}_k\rangle\langle\Psi^{N-1}_k|[\hat{H},\hat{O}_i]| \Psi^N_0\rangle}{\langle\Psi^N_0|\hat{O}_i^\dagger\hat{O}_i |\Psi^N_0\rangle}\nonumber\\
&=&-\sum_k\frac{\langle\Psi^N_0|\hat{O}_i^\dagger|\Psi^{N-1}_k\rangle\langle\Psi^{N-1}_k|\hat{O}_i| \Psi^N_0\rangle}{\langle\Psi^N_0|\hat{O}_i^\dagger\hat{O}_i |\Psi^N_0\rangle}(E^{N-1}_k-E^N_0)\nonumber\\
&=&\frac{\sum_kB^{k,R}_i\epsilon^R_k}{\sum_kB^{k,R}_i},
%\label{Eqn:valence-energy}
    \end{eqnarray}
 which is a weighted average of the poles of $G^R_{i}$. Inserting a complete set of eigenstates of the $(N+1)$-electron system in Eq.~(\ref{Eqn:conduction-energy}) one can show in a similar way that the  $\epsilon^A_i=\sum_kB^{k,A}_i\epsilon^A_k / \sum_kB^{k,A}_i$ within the EKT basis. This means that if there are not satellites in the EKT basis set, then the EKT removal/addition energies are exact, provided that one uses the exact first moment.

\subsubsection{Insights into the MEET}
\begin{figure}[t]
 \includegraphics[width=0.45\textwidth]{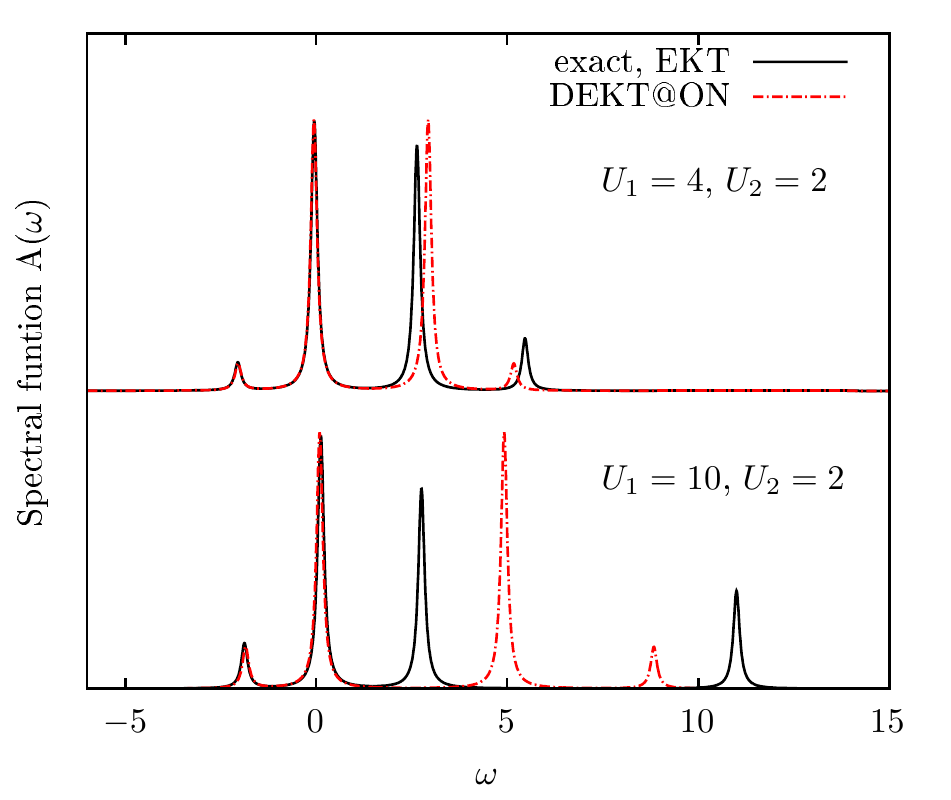}
 \caption{Spectral function of the Hubbard dimer with two different on-site interactions $U_1$ and $U_2$ for site 1 and site 2, respectively: exact solution (black curves) \textit{vs} EKT (black curves, EKT is exact in this case, see text) and DEKT/MEET in the basis of natural orbitals (red curves).}
 \label{Fig:Hubbard}
\end{figure}

More choices for an optimal basis set for the MEET expressions are now possible. In previous works we considered the basis of natural orbitals as optimal basis set based on the results obtained for the Hubbard model \cite{stefano}. However, in the Hubbard model the EKT and the MEET are the same since the $\Lambda^{R/A}$ matrices are diagonal in the basis of natural orbitals. Moreover, the EKT is exact for the symmetric Hubbard dimer, since the MEET is exact for this simple model at each level of approximation as discussed in Ref.~ \onlinecite{stefano}. Here we find that the EKT is exact also for the asymmetric case.  
This is clearly not the case in general. Therefore the basis which diagonalizes ($\Lambda^{R/A}$)
%or $\sqrt{n_in_j}\Lambda^R$ ($\sqrt{(1-n_i)(1-n_j)}\Lambda^A$) 
can be a better option. \Pina{We notice that this choice of the optimal basis set can be generally applied to other methods which express $G$ as a continued fraction, such as the Lanczos method \cite{Balzer_2011}, in order to have more accurate results at a given order of truncation of the series.}
For example in Fig.~\ref{Fig:Hubbard} we report the spectral function of the Hubbard dimer governed by the Hamiltonian in Eq.~(\ref{Eqn:Hubbard_H}) for two different values of $|U_1-U_2|$. The (D)EKT results are obtained using exact density matrices. The results show that the basis which diagonalizes the $\Lambda^{R/A}$ matrices is a much better choice than the basis of natural orbitals the more the difference $|U_1-U_2|$ is large. We also observe that the removal part is less affected by the diagonal approximation than the addition part, and we observe this trend also in more complex systems. The diagonal approximation has been addressed in literature also for realistic systems.\cite{Piris_TCA2013,kent_PRB1998} In particular in bulk silicon QMC results show that the DEKT slightly overestimates the EKT band gaps. Below we will address this system in more details.
%%%%%
%%%%%

 %%%%%
\subsection{Realistic systems: the example of bulk Si}
As an example of realistic systems we use bulk silicon, for which results using the EKT within QMC are reported in Ref.~\onlinecite{kent_PRB1998}. 
The diagonal approximation to the EKT within QMC works very well for the valence states and slightly less well for the conduction states, with a band gap at the $\Gamma$ point of 4.4 eV \textit{vs} 3.8 eV from the full EKT compared to 3.4 eV in experiment.\cite{kent_PRB1998}
We note that this discrepency is largly due to the energy of the conduction band at $\Gamma$. 
However, bulk silicon is a relatively weakly correlated system, with hence a predominant quasiparticle-like spectral function for which the EKT is a good approximation. Larger overestimation of the band gap can be expected for strongly correlated systems. This can be understood from the interpretation of the EKT energies as first moments of the one-body Green's function. However, an important point to stress is that even for these systems, which are a challenge for  state-of-the-art \textit{ab initio} methods, such as GW,  the EKT would open a gap, in accordance with experiment. 

As pointed out in Ref.\onlinecite{kent_PRB1998} the choice of the trial wave function for QMC calculations
is of critical importance. Indeed, as a result of the fixed-node approximation, QMC calculations of the matrix elements of the density matrix and operators $\mathbf{V}^{R,A}$
[see EKT equations (\ref{Eqn:removal-energy}) and (\ref{Eqn:addition-energy})] are expected to critically depend
on the nodal structure of the trial wave function employed. For the weakly correlated bulk silicon the accurate QMC value of 3.8 eV reported above
has been obtained by Kent \textit{et al.} \cite{kent_PRB1998} using a standard Slater-Jastrow trial wave function whose nodes are those of a single determinant consisting of LDA orbitals.
For more strongly correlated systems the wave function acquires a significant multi-determinant character and getting physically meaningful nodes
becomes much more difficult. It is thus useful to use the EKT within alternative approaches.

\begin{figure}[t]
 \includegraphics[width=0.45\textwidth]{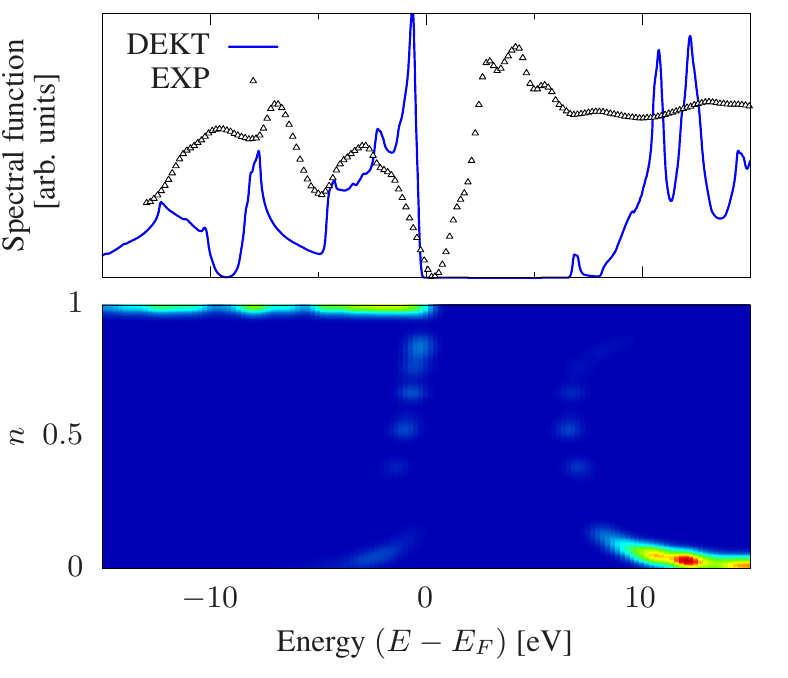}
 \caption{Spectral function of bulk Si within the DEKT. }
 \label{Fig:Si}
\end{figure}

\begin{figure}[t]
 \includegraphics[width=0.45\textwidth]{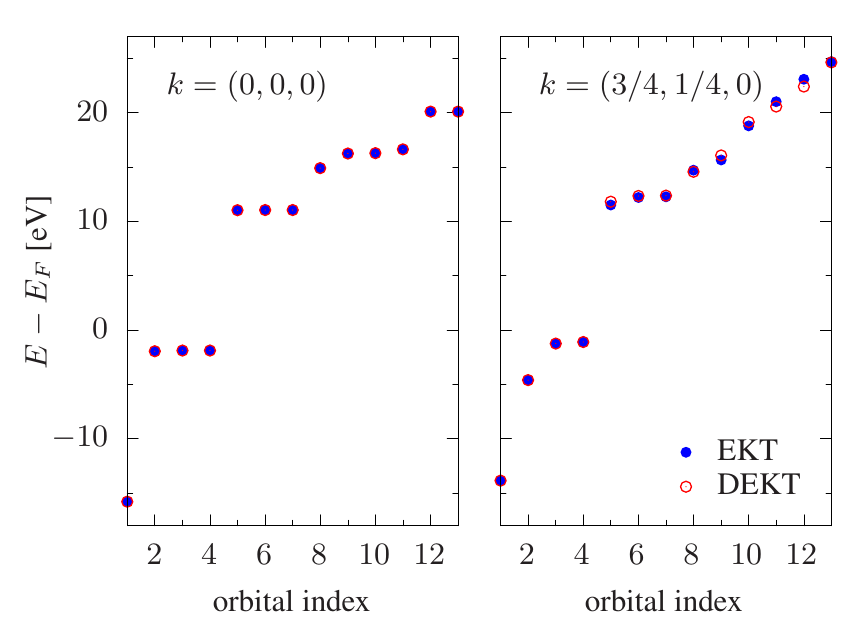}
 \caption{EKT vs DEKT for two $k$-points, $k=(0,0,0)$ and $k=(3/4,1/4,0)$.}
 \label{Fig:diago}
\end{figure}

In our previous works \cite{stefano,stefano_JCTC,stefano_PRR2021} we used reduced-density matrix functional theory (RDMFT) \cite{Gilbert75} to find approximations to the one- and two-body density matrices which are needed in the MEET equations. More specifically the two-body density matrix in the MEET equations is approximated using the Power functional, which is given by $\Gamma^{(2)}_{ijkl}=n_in_j\delta_{il}\delta_{jk}-n^\alpha_in^\alpha_j\delta_{ik}\delta_{jl}$ ($\alpha=0.65$).\cite{Sharma13}
The optimal natural orbitals $\{\phi_i\}$  and occupation numbers $\{n_i\}$  are obtained by minimizing the total energy which is expressed in terms of $\gamma$ and $\Gamma^{(2)}$, with $\Gamma^{(2)}$ as functional of $\gamma$. The Power functional is used also to approximate $\Gamma^{(2)}[\gamma]$ in the energy functional. In this work we use the same protocol for the EKT equations. We implemented the EKT equations in a modified version of the full-potential linearized augmented plane wave (FP-LAPW)
code ELK \cite{elk}, with practical details of the calculations
following the scheme described in Ref.~\onlinecite{sharma_PRB08}. For bulk Si we used a lattice constant of 5.43 \AA\,  and a $\Gamma$-centered $8\times8\times8$ ${\bf k}$-point sampling of the Brillouin zone.  
In Fig.~\ref{Fig:Si} we report the DEKT spectral function of bulk silicon: the direct band gap at $\Gamma$ is 12.9 eV, while the fundamental band gap is 8.18 eV, which is larger than the experimental one of $1.12$ eV \cite{Semiconductor_devices}. We also observe a spurious peak in the band gap due to the fact that the Power functional produces occupation numbers which strongly deviate from 1 and 0 (as one would expect for this weakly correlated system) close to the Fermi energy (see bottom panel of Fig.~\ref{Fig:Si}). This is in contrast with the QMC results, which points to a bad performance of the Power functional. Moreover, the full EKT does not show any improvement over the DEKT, as one can see from Fig.~\ref{Fig:diago}, in which the EKT and DEKT energies are reported:  the fundamental band gap is reduced by only 0.06 eV. This is again in contrast with the QMC results in which, although small, there is a significant  difference. We attribute this different trend to the use of the Power functional, which contracts the four-point 2-RDM to two points only, and hence probably mitigating the impact of the diagonalization of the $\Lambda^{R/A}$ matrices. These results on bulk Si indicate that, although the EKT/DEKT are expected to overestimate the band gap (even using very accurate density matrices), this overestimation can be much amplified by using approximations such as the Power functional. More advanced approximations to $\Gamma^{(2)}$ are hence needed, which give, in particular, more accurate natural occupation numbers. 
We notice that varying $\alpha$ would change the band gap width. In particular $\alpha=1$ would give the HF band gap, which still overestimates the experimental one, whereas decreasing $\alpha$ would increase the overestimation of the band gap.


%%%%%
\section{Conclusions and Perspectives\label{Sec:conclusions}}
We linked our recently derived Many-Body Effective Energy Theory (MEET) for the calculation of photoemission spectra to the Extended Koopman's Theorem (EKT). Within the lowest level of approximation in terms of one- and two-body density matrices, the MEET equations correspond to the so-called diagonal approximation to the EKT (DEKT) equations. This allowed us to readily extend the EKT to the calculation of an approximate spectral function as well as to give an alternative interpretation of the EKT in terms of moments of the one-body Green's function. Using the test case of the Hubbard dimer with two different on-site interactions $U_1$ and $U_2$ for site 1 and site 2 we showed the effect of the basis set on the MEET (removal and addition) energies: in particular HOMO-LUMO gap in the basis sets which solve the EKT secular equations (one basis set for the valence part and one for the conduction part) is smaller than the HOMO-LUMO gap obtained using the natural orbital basis set. These results are in line with the EKT results reported in literature for bulk Si using QMC. We have implemented the EKT within reduced-density matrix functional theory (RDMFT), which offers a convenient computationally affordable framework to treat extended systems. However one has to rely on approximate one- and two-body density matrices. We showed that using the currently available approximations the DEKT band gap of Si largely deviates from the DEKT value obtained using QMC (12.9 eV $vs$  4.4 eV at the $\Gamma$ point) and, moreover, there is no effect of the basis set (EKT \textit{vs} DEKT) on the DEKT energies, contrary to what is observed within QMC, where, although small, there is a significant difference.
These results on bulk Si indicate that, although the EKT/DEKT are expected to overestimate the band gap (even using very accurate density matrices), this overestimation can be much amplified by commonly used approximations in RDMFT. This also explains the huge overestimation of the band gap obtained by the MEET within RDMFT in strongly correlated systems such as paramagnetic NiO. We are currently working on improving approximations to correlation in RDMFT by introducing some form of screening (for example the screening due to electron-hole excitations as in GW), which is of particular importance in solids.

\begin{acknowledgments}
This study has been supported through the EUR grant NanoX ANR-17-EURE-0009 in the framework of the ``Programme des Investissements d'Avenir" and by ANR (project ANR-18-CE30-0025 and ANR-19-CE30-0011).
\end{acknowledgments}

\bibliography{bibliography}

\begin{thebibliography}{42}%
\makeatletter
\providecommand \@ifxundefined [1]{%
 \@ifx{#1\undefined}
}%
\providecommand \@ifnum [1]{%
 \ifnum #1\expandafter \@firstoftwo
 \else \expandafter \@secondoftwo
 \fi
}%
\providecommand \@ifx [1]{%
 \ifx #1\expandafter \@firstoftwo
 \else \expandafter \@secondoftwo
 \fi
}%
\providecommand \natexlab [1]{#1}%
\providecommand \enquote  [1]{``#1''}%
\providecommand \bibnamefont  [1]{#1}%
\providecommand \bibfnamefont [1]{#1}%
\providecommand \citenamefont [1]{#1}%
\providecommand \href@noop [0]{\@secondoftwo}%
\providecommand \href [0]{\begingroup \@sanitize@url \@href}%
\providecommand \@href[1]{\@@startlink{#1}\@@href}%
\providecommand \@@href[1]{\endgroup#1\@@endlink}%
\providecommand \@sanitize@url [0]{\catcode `\\12\catcode `\$12\catcode
  `\&12\catcode `\#12\catcode `\^12\catcode `\_12\catcode `\%12\relax}%
\providecommand \@@startlink[1]{}%
\providecommand \@@endlink[0]{}%
\providecommand \url  [0]{\begingroup\@sanitize@url \@url }%
\providecommand \@url [1]{\endgroup\@href {#1}{\urlprefix }}%
\providecommand \urlprefix  [0]{URL }%
\providecommand \Eprint [0]{\href }%
\providecommand \doibase [0]{http://dx.doi.org/}%
\providecommand \selectlanguage [0]{\@gobble}%
\providecommand \bibinfo  [0]{\@secondoftwo}%
\providecommand \bibfield  [0]{\@secondoftwo}%
\providecommand \translation [1]{[#1]}%
\providecommand \BibitemOpen [0]{}%
\providecommand \bibitemStop [0]{}%
\providecommand \bibitemNoStop [0]{.\EOS\space}%
\providecommand \EOS [0]{\spacefactor3000\relax}%
\providecommand \BibitemShut  [1]{\csname bibitem#1\endcsname}%
\let\auto@bib@innerbib\@empty
%</preamble>
\bibitem [{\citenamefont {Morrell}\ \emph {et~al.}(1975)\citenamefont
  {Morrell}, \citenamefont {Parr},\ and\ \citenamefont
  {Levy}}]{morrell_JCP1975}%
  \BibitemOpen
  \bibfield  {author} {\bibinfo {author} {\bibfnamefont {Marilyn~M.}\
  \bibnamefont {Morrell}}, \bibinfo {author} {\bibfnamefont {Robert~G.}\
  \bibnamefont {Parr}}, \ and\ \bibinfo {author} {\bibfnamefont {Mel}\
  \bibnamefont {Levy}},\ }\bibfield  {title} {\enquote {\bibinfo {title}
  {Calculation of ionization potentials from density matrices and natural
  functions, and the long‐range behavior of natural orbitals and electron
  density},}\ }\href {\doibase 10.1063/1.430509} {\bibfield  {journal}
  {\bibinfo  {journal} {The Journal of Chemical Physics}\ }\textbf {\bibinfo
  {volume} {62}},\ \bibinfo {pages} {549--554} (\bibinfo {year} {1975})},\
  \Eprint
  {http://arxiv.org/abs/https://aip.scitation.org/doi/pdf/10.1063/1.430509}
  {https://aip.scitation.org/doi/pdf/10.1063/1.430509} \BibitemShut {NoStop}%
\bibitem [{\citenamefont {Smith}\ and\ \citenamefont
  {Day}(1975)}]{smith_JCP1975}%
  \BibitemOpen
  \bibfield  {author} {\bibinfo {author} {\bibfnamefont {Darwin~W.}\
  \bibnamefont {Smith}}\ and\ \bibinfo {author} {\bibfnamefont {Orville~W.}\
  \bibnamefont {Day}},\ }\bibfield  {title} {\enquote {\bibinfo {title}
  {Extension of koopmans’ theorem. i. derivation},}\ }\href {\doibase
  10.1063/1.430253} {\bibfield  {journal} {\bibinfo  {journal} {The Journal of
  Chemical Physics}\ }\textbf {\bibinfo {volume} {62}},\ \bibinfo {pages}
  {113--114} (\bibinfo {year} {1975})},\ \Eprint
  {http://arxiv.org/abs/https://aip.scitation.org/doi/pdf/10.1063/1.430253}
  {https://aip.scitation.org/doi/pdf/10.1063/1.430253} \BibitemShut {NoStop}%
\bibitem [{\citenamefont {Gilbert}(1975{\natexlab{a}})}]{Gilbert}%
  \BibitemOpen
  \bibfield  {author} {\bibinfo {author} {\bibfnamefont {T.~L.}\ \bibnamefont
  {Gilbert}},\ }\bibfield  {title} {\enquote {\bibinfo {title} {Hohenberg-kohn
  theorem for nonlocal external potentials},}\ }\href {\doibase
  10.1103/PhysRevB.12.2111} {\bibfield  {journal} {\bibinfo  {journal} {Phys.
  Rev. B}\ }\textbf {\bibinfo {volume} {12}},\ \bibinfo {pages} {2111--2120}
  (\bibinfo {year} {1975}{\natexlab{a}})}\BibitemShut {NoStop}%
\bibitem [{\citenamefont {Hedin}(1965)}]{Hedin65}%
  \BibitemOpen
  \bibfield  {author} {\bibinfo {author} {\bibfnamefont {Lars}\ \bibnamefont
  {Hedin}},\ }\bibfield  {title} {\enquote {\bibinfo {title} {New method for
  calculating the one-particle green's function with application to the
  electron-gas problem},}\ }\href@noop {} {\bibfield  {journal} {\bibinfo
  {journal} {Phys. Rev.}\ }\textbf {\bibinfo {volume} {139}},\ \bibinfo {pages}
  {A796--A823} (\bibinfo {year} {1965})}\BibitemShut {NoStop}%
\bibitem [{\citenamefont {Pernal}\ and\ \citenamefont
  {Cioslowski}(2005)}]{pernal_CPL2005}%
  \BibitemOpen
  \bibfield  {author} {\bibinfo {author} {\bibfnamefont {Katarzyna}\
  \bibnamefont {Pernal}}\ and\ \bibinfo {author} {\bibfnamefont {Jerzy}\
  \bibnamefont {Cioslowski}},\ }\bibfield  {title} {\enquote {\bibinfo {title}
  {Ionization potentials from the extended koopmans’ theorem applied to
  density matrix functional theory},}\ }\href {\doibase
  https://doi.org/10.1016/j.cplett.2005.06.103} {\bibfield  {journal} {\bibinfo
   {journal} {Chemical Physics Letters}\ }\textbf {\bibinfo {volume} {412}},\
  \bibinfo {pages} {71--75} (\bibinfo {year} {2005})}\BibitemShut {NoStop}%
\bibitem [{\citenamefont {Leiva}\ and\ \citenamefont
  {Piris}(2005)}]{piris_JCP2005}%
  \BibitemOpen
  \bibfield  {author} {\bibinfo {author} {\bibfnamefont {P.}~\bibnamefont
  {Leiva}}\ and\ \bibinfo {author} {\bibfnamefont {M.}~\bibnamefont {Piris}},\
  }\bibfield  {title} {\enquote {\bibinfo {title} {Assessment of a new approach
  for the two-electron cumulant in natural-orbital-functional theory},}\ }\href
  {\doibase 10.1063/1.2135289} {\bibfield  {journal} {\bibinfo  {journal} {The
  Journal of Chemical Physics}\ }\textbf {\bibinfo {volume} {123}},\ \bibinfo
  {pages} {214102} (\bibinfo {year} {2005})},\ \Eprint
  {http://arxiv.org/abs/https://doi.org/10.1063/1.2135289}
  {https://doi.org/10.1063/1.2135289} \BibitemShut {NoStop}%
\bibitem [{\citenamefont {Piris}\ \emph {et~al.}(2012)\citenamefont {Piris},
  \citenamefont {Matxain}, \citenamefont {Lopez},\ and\ \citenamefont
  {Ugalde}}]{piris_JCP2012}%
  \BibitemOpen
  \bibfield  {author} {\bibinfo {author} {\bibfnamefont {Mario}\ \bibnamefont
  {Piris}}, \bibinfo {author} {\bibfnamefont {Jon~M.}\ \bibnamefont {Matxain}},
  \bibinfo {author} {\bibfnamefont {Xabier}\ \bibnamefont {Lopez}}, \ and\
  \bibinfo {author} {\bibfnamefont {Jesus~M.}\ \bibnamefont {Ugalde}},\
  }\bibfield  {title} {\enquote {\bibinfo {title} {The extended koopmans’
  theorem: Vertical ionization potentials from natural orbital functional
  theory},}\ }\href {\doibase 10.1063/1.4709769} {\bibfield  {journal}
  {\bibinfo  {journal} {The Journal of Chemical Physics}\ }\textbf {\bibinfo
  {volume} {136}},\ \bibinfo {pages} {174116} (\bibinfo {year} {2012})},\
  \Eprint {http://arxiv.org/abs/https://doi.org/10.1063/1.4709769}
  {https://doi.org/10.1063/1.4709769} \BibitemShut {NoStop}%
\bibitem [{\citenamefont {Piris}\ \emph {et~al.}(2013)\citenamefont {Piris},
  \citenamefont {Matxain}, \citenamefont {Lopez},\ and\ \citenamefont
  {Ugalde}}]{Piris_TCA2013}%
  \BibitemOpen
  \bibfield  {author} {\bibinfo {author} {\bibfnamefont {Mario}\ \bibnamefont
  {Piris}}, \bibinfo {author} {\bibfnamefont {Jon~M.}\ \bibnamefont {Matxain}},
  \bibinfo {author} {\bibfnamefont {Xabier}\ \bibnamefont {Lopez}}, \ and\
  \bibinfo {author} {\bibfnamefont {Jesus~M.}\ \bibnamefont {Ugalde}},\
  }\bibfield  {title} {\enquote {\bibinfo {title} {The one-electron picture in
  the piris natural orbital functional 5 (pnof5)},}\ }\href {\doibase
  10.1007/s00214-012-1298-4} {\bibfield  {journal} {\bibinfo  {journal}
  {Theoretical Chemistry Accounts}\ }\textbf {\bibinfo {volume} {132}},\
  \bibinfo {pages} {1298} (\bibinfo {year} {2013})}\BibitemShut {NoStop}%
\bibitem [{\citenamefont {Dahlen}\ and\ \citenamefont {van
  Leeuwen}(2005)}]{dahlen_JCP2005}%
  \BibitemOpen
  \bibfield  {author} {\bibinfo {author} {\bibfnamefont {Nils~Erik}\
  \bibnamefont {Dahlen}}\ and\ \bibinfo {author} {\bibfnamefont {Robert}\
  \bibnamefont {van Leeuwen}},\ }\bibfield  {title} {\enquote {\bibinfo {title}
  {Self-consistent solution of the dyson equation for atoms and molecules
  within a conserving approximation},}\ }\href {\doibase 10.1063/1.1884965}
  {\bibfield  {journal} {\bibinfo  {journal} {The Journal of Chemical Physics}\
  }\textbf {\bibinfo {volume} {122}},\ \bibinfo {pages} {164102} (\bibinfo
  {year} {2005})},\ \Eprint
  {http://arxiv.org/abs/https://doi.org/10.1063/1.1884965}
  {https://doi.org/10.1063/1.1884965} \BibitemShut {NoStop}%
\bibitem [{\citenamefont {Stan}\ \emph {et~al.}(2006)\citenamefont {Stan},
  \citenamefont {Dahlen},\ and\ \citenamefont {van Leeuwen}}]{stan_EPL2006}%
  \BibitemOpen
  \bibfield  {author} {\bibinfo {author} {\bibfnamefont {A}~\bibnamefont
  {Stan}}, \bibinfo {author} {\bibfnamefont {N.~E}\ \bibnamefont {Dahlen}}, \
  and\ \bibinfo {author} {\bibfnamefont {R.}~\bibnamefont {van Leeuwen}},\
  }\bibfield  {title} {\enquote {\bibinfo {title} {Fully self-consistent {GW}
  calculations for atoms and molecules},}\ }\href {\doibase
  10.1209/epl/i2006-10266-6} {\bibfield  {journal} {\bibinfo  {journal}
  {Europhysics Letters ({EPL})}\ }\textbf {\bibinfo {volume} {76}},\ \bibinfo
  {pages} {298--304} (\bibinfo {year} {2006})}\BibitemShut {NoStop}%
\bibitem [{\citenamefont {Stan}\ \emph {et~al.}(2009)\citenamefont {Stan},
  \citenamefont {Dahlen},\ and\ \citenamefont {van Leeuwen}}]{stan_JCP2009}%
  \BibitemOpen
  \bibfield  {author} {\bibinfo {author} {\bibfnamefont {Adrian}\ \bibnamefont
  {Stan}}, \bibinfo {author} {\bibfnamefont {Nils~Erik}\ \bibnamefont
  {Dahlen}}, \ and\ \bibinfo {author} {\bibfnamefont {Robert}\ \bibnamefont
  {van Leeuwen}},\ }\bibfield  {title} {\enquote {\bibinfo {title} {Levels of
  self-consistency in the gw approximation},}\ }\href {\doibase
  10.1063/1.3089567} {\bibfield  {journal} {\bibinfo  {journal} {The Journal of
  Chemical Physics}\ }\textbf {\bibinfo {volume} {130}},\ \bibinfo {pages}
  {114105} (\bibinfo {year} {2009})},\ \Eprint
  {http://arxiv.org/abs/https://doi.org/10.1063/1.3089567}
  {https://doi.org/10.1063/1.3089567} \BibitemShut {NoStop}%
\bibitem [{\citenamefont {Kent}\ \emph {et~al.}(1998)\citenamefont {Kent},
  \citenamefont {Hood}, \citenamefont {Towler}, \citenamefont {Needs},\ and\
  \citenamefont {Rajagopal}}]{kent_PRB1998}%
  \BibitemOpen
  \bibfield  {author} {\bibinfo {author} {\bibfnamefont {P.~R.~C.}\
  \bibnamefont {Kent}}, \bibinfo {author} {\bibfnamefont {Randolph~Q.}\
  \bibnamefont {Hood}}, \bibinfo {author} {\bibfnamefont {M.~D.}\ \bibnamefont
  {Towler}}, \bibinfo {author} {\bibfnamefont {R.~J.}\ \bibnamefont {Needs}}, \
  and\ \bibinfo {author} {\bibfnamefont {G.}~\bibnamefont {Rajagopal}},\
  }\bibfield  {title} {\enquote {\bibinfo {title} {Quantum monte carlo
  calculations of the one-body density matrix and excitation energies of
  silicon},}\ }\href {\doibase 10.1103/PhysRevB.57.15293} {\bibfield  {journal}
  {\bibinfo  {journal} {Phys. Rev. B}\ }\textbf {\bibinfo {volume} {57}},\
  \bibinfo {pages} {15293--15302} (\bibinfo {year} {1998})}\BibitemShut
  {NoStop}%
\bibitem [{\citenamefont {Zheng}(2016)}]{zhengPhD_2016}%
  \BibitemOpen
  \bibfield  {author} {\bibinfo {author} {\bibfnamefont {Huihuo}\ \bibnamefont
  {Zheng}},\ }\emph {\bibinfo {title} {First principles quantum Monte Carlo
  study of correlated electronic systems}},\ \href
  {http://hdl.handle.net/2142/92794} {Ph.D. thesis},\ \bibinfo  {school}
  {University of Illinois at Urbana-Champaign} (\bibinfo {year}
  {2016})\BibitemShut {NoStop}%
\bibitem [{\citenamefont {Cioslowski}\ \emph {et~al.}(1997)\citenamefont
  {Cioslowski}, \citenamefont {Piskorz},\ and\ \citenamefont
  {Liu}}]{cioslowsk_JCP1997}%
  \BibitemOpen
  \bibfield  {author} {\bibinfo {author} {\bibfnamefont {Jerzy}\ \bibnamefont
  {Cioslowski}}, \bibinfo {author} {\bibfnamefont {Pawel}\ \bibnamefont
  {Piskorz}}, \ and\ \bibinfo {author} {\bibfnamefont {Guanghua}\ \bibnamefont
  {Liu}},\ }\bibfield  {title} {\enquote {\bibinfo {title} {Ionization
  potentials and electron affinities from the extended koopmans’ theorem
  applied to energy-derivative density matrices: The ektmpn and ektqcisd
  methods},}\ }\href {\doibase 10.1063/1.474921} {\bibfield  {journal}
  {\bibinfo  {journal} {The Journal of Chemical Physics}\ }\textbf {\bibinfo
  {volume} {107}},\ \bibinfo {pages} {6804--6811} (\bibinfo {year} {1997})},\
  \Eprint {http://arxiv.org/abs/https://doi.org/10.1063/1.474921}
  {https://doi.org/10.1063/1.474921} \BibitemShut {NoStop}%
\bibitem [{\citenamefont {Bozkaya}(2013)}]{bozkaya_JCP2013}%
  \BibitemOpen
  \bibfield  {author} {\bibinfo {author} {\bibfnamefont {Uğur}\ \bibnamefont
  {Bozkaya}},\ }\bibfield  {title} {\enquote {\bibinfo {title} {The extended
  koopmans' theorem for orbital-optimized methods: Accurate computation of
  ionization potentials},}\ }\href {\doibase 10.1063/1.4825041} {\bibfield
  {journal} {\bibinfo  {journal} {The Journal of Chemical Physics}\ }\textbf
  {\bibinfo {volume} {139}},\ \bibinfo {pages} {154105} (\bibinfo {year}
  {2013})},\ \Eprint {http://arxiv.org/abs/https://doi.org/10.1063/1.4825041}
  {https://doi.org/10.1063/1.4825041} \BibitemShut {NoStop}%
\bibitem [{\citenamefont {Bozkaya}\ and\ \citenamefont
  {Ünal}(2018)}]{bozkaya_JPCA2018}%
  \BibitemOpen
  \bibfield  {author} {\bibinfo {author} {\bibfnamefont {Uğur}\ \bibnamefont
  {Bozkaya}}\ and\ \bibinfo {author} {\bibfnamefont {Aslı}\ \bibnamefont
  {Ünal}},\ }\bibfield  {title} {\enquote {\bibinfo {title} {State-of-the-art
  computations of vertical ionization potentials with the extended koopmans’
  theorem integrated with the ccsd(t) method},}\ }\href {\doibase
  10.1021/acs.jpca.8b01851} {\bibfield  {journal} {\bibinfo  {journal} {The
  Journal of Physical Chemistry A}\ }\textbf {\bibinfo {volume} {122}},\
  \bibinfo {pages} {4375--4380} (\bibinfo {year} {2018})},\ \bibinfo {note}
  {pMID: 29517234},\ \Eprint
  {http://arxiv.org/abs/https://doi.org/10.1021/acs.jpca.8b01851}
  {https://doi.org/10.1021/acs.jpca.8b01851} \BibitemShut {NoStop}%
\bibitem [{\citenamefont {Pavlyukh}(2019)}]{Pavlyukh_PSS2019}%
  \BibitemOpen
  \bibfield  {author} {\bibinfo {author} {\bibfnamefont {Yaroslav}\
  \bibnamefont {Pavlyukh}},\ }\bibfield  {title} {\enquote {\bibinfo {title}
  {The ubiquitous extended koopmans‚Äô theorem},}\ }\href {\doibase
  https://doi.org/10.1002/pssb.201800591} {\bibfield  {journal} {\bibinfo
  {journal} {physica status solidi (b)}\ }\textbf {\bibinfo {volume} {256}},\
  \bibinfo {pages} {1800591} (\bibinfo {year} {2019})},\ \Eprint
  {http://arxiv.org/abs/https://onlinelibrary.wiley.com/doi/pdf/10.1002/pssb.201800591}
  {https://onlinelibrary.wiley.com/doi/pdf/10.1002/pssb.201800591} \BibitemShut
  {NoStop}%
\bibitem [{\citenamefont {Katriel}\ and\ \citenamefont
  {Davidson}(1980)}]{katriel_pnas1980}%
  \BibitemOpen
  \bibfield  {author} {\bibinfo {author} {\bibfnamefont {J}~\bibnamefont
  {Katriel}}\ and\ \bibinfo {author} {\bibfnamefont {E.R.}\ \bibnamefont
  {Davidson}},\ }\bibfield  {title} {\enquote {\bibinfo {title} {Asymptotic
  behavior of atomic and molecular wave functions},}\ }\href {\doibase
  10.1073/pnas.77.8.4403} {\bibfield  {journal} {\bibinfo  {journal}
  {Proceedings of the National Academy of Sciences of the United States of
  America}\ }\textbf {\bibinfo {volume} {77}},\ \bibinfo {pages} {4403--6}
  (\bibinfo {year} {1980})}\BibitemShut {NoStop}%
\bibitem [{\citenamefont {Sundholm}\ and\ \citenamefont
  {Olsen}(1993)}]{olsen_JCP1993}%
  \BibitemOpen
  \bibfield  {author} {\bibinfo {author} {\bibfnamefont {Dage}\ \bibnamefont
  {Sundholm}}\ and\ \bibinfo {author} {\bibfnamefont {Jeppe}\ \bibnamefont
  {Olsen}},\ }\bibfield  {title} {\enquote {\bibinfo {title} {The exactness of
  the extended koopmans’ theorem: A numerical study},}\ }\href {\doibase
  10.1063/1.464028} {\bibfield  {journal} {\bibinfo  {journal} {The Journal of
  Chemical Physics}\ }\textbf {\bibinfo {volume} {98}},\ \bibinfo {pages}
  {3999--4002} (\bibinfo {year} {1993})},\ \Eprint
  {http://arxiv.org/abs/https://doi.org/10.1063/1.464028}
  {https://doi.org/10.1063/1.464028} \BibitemShut {NoStop}%
\bibitem [{\citenamefont {Romaniello}\ \emph {et~al.}(2009)\citenamefont
  {Romaniello}, \citenamefont {Guyot},\ and\ \citenamefont
  {Reining}}]{romaniello2009}%
  \BibitemOpen
  \bibfield  {author} {\bibinfo {author} {\bibfnamefont {P.}~\bibnamefont
  {Romaniello}}, \bibinfo {author} {\bibfnamefont {S.}~\bibnamefont {Guyot}}, \
  and\ \bibinfo {author} {\bibfnamefont {L.}~\bibnamefont {Reining}},\
  }\bibfield  {title} {\enquote {\bibinfo {title} {The self-energy beyond gw:
  Local and nonlocal vertex corrections},}\ }\href {\doibase 10.1063/1.3249965}
  {\bibfield  {journal} {\bibinfo  {journal} {J. Chem. Phys.}\ }\textbf
  {\bibinfo {volume} {131}},\ \bibinfo {pages} {154111} (\bibinfo {year}
  {2009})},\ \Eprint {http://arxiv.org/abs/https://doi.org/10.1063/1.3249965}
  {https://doi.org/10.1063/1.3249965} \BibitemShut {NoStop}%
\bibitem [{\citenamefont {Romaniello}\ \emph {et~al.}(2012)\citenamefont
  {Romaniello}, \citenamefont {Bechstedt},\ and\ \citenamefont
  {Reining}}]{romaniello2012}%
  \BibitemOpen
  \bibfield  {author} {\bibinfo {author} {\bibfnamefont {Pina}\ \bibnamefont
  {Romaniello}}, \bibinfo {author} {\bibfnamefont {Friedhelm}\ \bibnamefont
  {Bechstedt}}, \ and\ \bibinfo {author} {\bibfnamefont {Lucia}\ \bibnamefont
  {Reining}},\ }\bibfield  {title} {\enquote {\bibinfo {title} {Beyond the $gw$
  approximation: Combining correlation channels},}\ }\href {\doibase
  10.1103/PhysRevB.85.155131} {\bibfield  {journal} {\bibinfo  {journal} {Phys.
  Rev. B}\ }\textbf {\bibinfo {volume} {85}},\ \bibinfo {pages} {155131}
  (\bibinfo {year} {2012})}\BibitemShut {NoStop}%
\bibitem [{\citenamefont {{Di Sabatino}}\ \emph {et~al.}(2015)\citenamefont
  {{Di Sabatino}}, \citenamefont {Berger}, \citenamefont {Reining},\ and\
  \citenamefont {Romaniello}}]{stefano_JCP2015}%
  \BibitemOpen
  \bibfield  {author} {\bibinfo {author} {\bibfnamefont {S.}~\bibnamefont {{Di
  Sabatino}}}, \bibinfo {author} {\bibfnamefont {J.~A.}\ \bibnamefont
  {Berger}}, \bibinfo {author} {\bibfnamefont {L.}~\bibnamefont {Reining}}, \
  and\ \bibinfo {author} {\bibfnamefont {P.}~\bibnamefont {Romaniello}},\
  }\bibfield  {title} {\enquote {\bibinfo {title} {{Reduced density-matrix
  functional theory: Correlation and spectroscopy}},}\ }\href {\doibase
  10.1063/1.4926327} {\bibfield  {journal} {\bibinfo  {journal} {J. Chem.
  Phys}\ }\textbf {\bibinfo {volume} {143}},\ \bibinfo {eid} {024108} (\bibinfo
  {year} {2015})}\BibitemShut {NoStop}%
\bibitem [{\citenamefont {Di~Sabatino}\ \emph {et~al.}(2016)\citenamefont
  {Di~Sabatino}, \citenamefont {Berger}, \citenamefont {Reining},\ and\
  \citenamefont {Romaniello}}]{stefano}%
  \BibitemOpen
  \bibfield  {author} {\bibinfo {author} {\bibfnamefont {Stefano}\ \bibnamefont
  {Di~Sabatino}}, \bibinfo {author} {\bibfnamefont {J.~A.}\ \bibnamefont
  {Berger}}, \bibinfo {author} {\bibfnamefont {Lucia}\ \bibnamefont {Reining}},
  \ and\ \bibinfo {author} {\bibfnamefont {Pina}\ \bibnamefont {Romaniello}},\
  }\bibfield  {title} {\enquote {\bibinfo {title} {Photoemission spectra from
  reduced density matrices: The band gap in strongly correlated systems},}\
  }\href {\doibase 10.1103/PhysRevB.94.155141} {\bibfield  {journal} {\bibinfo
  {journal} {Phys. Rev. B}\ }\textbf {\bibinfo {volume} {94}},\ \bibinfo
  {pages} {155141} (\bibinfo {year} {2016})}\BibitemShut {NoStop}%
\bibitem [{\citenamefont {Di~Sabatino}\ \emph {et~al.}(2021)\citenamefont
  {Di~Sabatino}, \citenamefont {Koskelo}, \citenamefont {Berger},\ and\
  \citenamefont {Romaniello}}]{stefano_PRR2021}%
  \BibitemOpen
  \bibfield  {author} {\bibinfo {author} {\bibfnamefont {S.}~\bibnamefont
  {Di~Sabatino}}, \bibinfo {author} {\bibfnamefont {J.}~\bibnamefont
  {Koskelo}}, \bibinfo {author} {\bibfnamefont {J.~A.}\ \bibnamefont {Berger}},
  \ and\ \bibinfo {author} {\bibfnamefont {P.}~\bibnamefont {Romaniello}},\
  }\bibfield  {title} {\enquote {\bibinfo {title} {Photoemission spectrum in
  paramagnetic feo under pressure: Towards an ab initio description},}\ }\href
  {\doibase 10.1103/PhysRevResearch.3.013172} {\bibfield  {journal} {\bibinfo
  {journal} {Phys. Rev. Research}\ }\textbf {\bibinfo {volume} {3}},\ \bibinfo
  {pages} {013172} (\bibinfo {year} {2021})}\BibitemShut {NoStop}%
\bibitem [{\citenamefont {Springer}\ \emph {et~al.}(1998)\citenamefont
  {Springer}, \citenamefont {Aryasetiawan},\ and\ \citenamefont
  {Karlsson}}]{Springer1998}%
  \BibitemOpen
  \bibfield  {author} {\bibinfo {author} {\bibfnamefont {M.}~\bibnamefont
  {Springer}}, \bibinfo {author} {\bibfnamefont {F.}~\bibnamefont
  {Aryasetiawan}}, \ and\ \bibinfo {author} {\bibfnamefont {K.}~\bibnamefont
  {Karlsson}},\ }\bibfield  {title} {\enquote {\bibinfo {title}
  {First-principles $\mathit{T}$-matrix theory with application to the 6 ev
  satellite in ni},}\ }\href {\doibase 10.1103/PhysRevLett.80.2389} {\bibfield
  {journal} {\bibinfo  {journal} {Phys. Rev. Lett.}\ }\textbf {\bibinfo
  {volume} {80}},\ \bibinfo {pages} {2389--2392} (\bibinfo {year}
  {1998})}\BibitemShut {NoStop}%
\bibitem [{\citenamefont {Zhukov}\ \emph {et~al.}(2004)\citenamefont {Zhukov},
  \citenamefont {Chulkov},\ and\ \citenamefont {Echenique}}]{Chulkov_PRL}%
  \BibitemOpen
  \bibfield  {author} {\bibinfo {author} {\bibfnamefont {V.~P.}\ \bibnamefont
  {Zhukov}}, \bibinfo {author} {\bibfnamefont {E.~V.}\ \bibnamefont {Chulkov}},
  \ and\ \bibinfo {author} {\bibfnamefont {P.~M.}\ \bibnamefont {Echenique}},\
  }\bibfield  {title} {\enquote {\bibinfo {title} {Lifetimes of excited
  electrons in fe and ni: First-principles gw and the $t$-matrix theory},}\
  }\href {\doibase 10.1103/PhysRevLett.93.096401} {\bibfield  {journal}
  {\bibinfo  {journal} {Phys. Rev. Lett.}\ }\textbf {\bibinfo {volume} {93}},\
  \bibinfo {pages} {096401} (\bibinfo {year} {2004})}\BibitemShut {NoStop}%
\bibitem [{\citenamefont {Guzzo}\ \emph {et~al.}(2011)\citenamefont {Guzzo},
  \citenamefont {Lani}, \citenamefont {Sottile}, \citenamefont {Romaniello},
  \citenamefont {Gatti}, \citenamefont {Kas}, \citenamefont {Rehr},
  \citenamefont {Silly}, \citenamefont {Sirotti},\ and\ \citenamefont
  {Reining}}]{guzzo_prl}%
  \BibitemOpen
  \bibfield  {author} {\bibinfo {author} {\bibfnamefont {Matteo}\ \bibnamefont
  {Guzzo}}, \bibinfo {author} {\bibfnamefont {Giovanna}\ \bibnamefont {Lani}},
  \bibinfo {author} {\bibfnamefont {Francesco}\ \bibnamefont {Sottile}},
  \bibinfo {author} {\bibfnamefont {Pina}\ \bibnamefont {Romaniello}}, \bibinfo
  {author} {\bibfnamefont {Matteo}\ \bibnamefont {Gatti}}, \bibinfo {author}
  {\bibfnamefont {Joshua~J.}\ \bibnamefont {Kas}}, \bibinfo {author}
  {\bibfnamefont {John~J.}\ \bibnamefont {Rehr}}, \bibinfo {author}
  {\bibfnamefont {Mathieu~G.}\ \bibnamefont {Silly}}, \bibinfo {author}
  {\bibfnamefont {Fausto}\ \bibnamefont {Sirotti}}, \ and\ \bibinfo {author}
  {\bibfnamefont {Lucia}\ \bibnamefont {Reining}},\ }\bibfield  {title}
  {\enquote {\bibinfo {title} {Valence electron photoemission spectrum of
  semiconductors: \textit{Ab~Initio} description of multiple satellites},}\
  }\href {\doibase 10.1103/PhysRevLett.107.166401} {\bibfield  {journal}
  {\bibinfo  {journal} {Phys. Rev. Lett.}\ }\textbf {\bibinfo {volume} {107}},\
  \bibinfo {pages} {166401} (\bibinfo {year} {2011})}\BibitemShut {NoStop}%
\bibitem [{\citenamefont {Lischner}\ \emph {et~al.}(2013)\citenamefont
  {Lischner}, \citenamefont {Vigil-Fowler},\ and\ \citenamefont
  {Louie}}]{louie2013}%
  \BibitemOpen
  \bibfield  {author} {\bibinfo {author} {\bibfnamefont {Johannes}\
  \bibnamefont {Lischner}}, \bibinfo {author} {\bibfnamefont {Derek}\
  \bibnamefont {Vigil-Fowler}}, \ and\ \bibinfo {author} {\bibfnamefont
  {Steven~G.}\ \bibnamefont {Louie}},\ }\bibfield  {title} {\enquote {\bibinfo
  {title} {Physical origin of satellites in photoemission of doped graphene: An
  ab initio $gw$ plus cumulant study},}\ }\href {\doibase
  10.1103/PhysRevLett.110.146801} {\bibfield  {journal} {\bibinfo  {journal}
  {Phys. Rev. Lett.}\ }\textbf {\bibinfo {volume} {110}},\ \bibinfo {pages}
  {146801} (\bibinfo {year} {2013})}\BibitemShut {NoStop}%
\bibitem [{\citenamefont {Shishkin}\ \emph {et~al.}(2007)\citenamefont
  {Shishkin}, \citenamefont {Marsman},\ and\ \citenamefont
  {Kresse}}]{Kresse_PRL07}%
  \BibitemOpen
  \bibfield  {author} {\bibinfo {author} {\bibfnamefont {M.}~\bibnamefont
  {Shishkin}}, \bibinfo {author} {\bibfnamefont {M.}~\bibnamefont {Marsman}}, \
  and\ \bibinfo {author} {\bibfnamefont {G.}~\bibnamefont {Kresse}},\
  }\bibfield  {title} {\enquote {\bibinfo {title} {Accurate quasiparticle
  spectra from self-consistent \textit{GW} calculations with vertex
  corrections},}\ }\href {\doibase 10.1103/PhysRevLett.99.246403} {\bibfield
  {journal} {\bibinfo  {journal} {Phys. Rev. Lett.}\ }\textbf {\bibinfo
  {volume} {99}},\ \bibinfo {pages} {246403} (\bibinfo {year}
  {2007})}\BibitemShut {NoStop}%
\bibitem [{\citenamefont {Kune\ifmmode~\check{s}\else \v{s}\fi{}}\ \emph
  {et~al.}(2007)\citenamefont {Kune\ifmmode~\check{s}\else \v{s}\fi{}},
  \citenamefont {Anisimov}, \citenamefont {Skornyakov}, \citenamefont
  {Lukoyanov},\ and\ \citenamefont {Vollhardt}}]{Vollhardt2007}%
  \BibitemOpen
  \bibfield  {author} {\bibinfo {author} {\bibfnamefont {J.}~\bibnamefont
  {Kune\ifmmode~\check{s}\else \v{s}\fi{}}}, \bibinfo {author} {\bibfnamefont
  {V.~I.}\ \bibnamefont {Anisimov}}, \bibinfo {author} {\bibfnamefont {S.~L.}\
  \bibnamefont {Skornyakov}}, \bibinfo {author} {\bibfnamefont {A.~V.}\
  \bibnamefont {Lukoyanov}}, \ and\ \bibinfo {author} {\bibfnamefont
  {D.}~\bibnamefont {Vollhardt}},\ }\bibfield  {title} {\enquote {\bibinfo
  {title} {Nio: Correlated band structure of a charge-transfer insulator},}\
  }\href {\doibase 10.1103/PhysRevLett.99.156404} {\bibfield  {journal}
  {\bibinfo  {journal} {Phys. Rev. Lett.}\ }\textbf {\bibinfo {volume} {99}},\
  \bibinfo {pages} {156404} (\bibinfo {year} {2007})}\BibitemShut {NoStop}%
\bibitem [{\citenamefont {Stefanucci}\ \emph {et~al.}(2014)\citenamefont
  {Stefanucci}, \citenamefont {Pavlyukh}, \citenamefont {Uimonen},\ and\
  \citenamefont {van Leeuwen}}]{stefanucci_PRB2014}%
  \BibitemOpen
  \bibfield  {author} {\bibinfo {author} {\bibfnamefont {G.}~\bibnamefont
  {Stefanucci}}, \bibinfo {author} {\bibfnamefont {Y.}~\bibnamefont
  {Pavlyukh}}, \bibinfo {author} {\bibfnamefont {A.-M.}\ \bibnamefont
  {Uimonen}}, \ and\ \bibinfo {author} {\bibfnamefont {R.}~\bibnamefont {van
  Leeuwen}},\ }\bibfield  {title} {\enquote {\bibinfo {title} {Diagrammatic
  expansion for positive spectral functions beyond $gw$: Application to vertex
  corrections in the electron gas},}\ }\href {\doibase
  10.1103/PhysRevB.90.115134} {\bibfield  {journal} {\bibinfo  {journal} {Phys.
  Rev. B}\ }\textbf {\bibinfo {volume} {90}},\ \bibinfo {pages} {115134}
  (\bibinfo {year} {2014})}\BibitemShut {NoStop}%
\bibitem [{\citenamefont {Lani}\ \emph {et~al.}(2012)\citenamefont {Lani},
  \citenamefont {Romaniello},\ and\ \citenamefont {Reining}}]{Lani_2012}%
  \BibitemOpen
  \bibfield  {author} {\bibinfo {author} {\bibfnamefont {Giovanna}\
  \bibnamefont {Lani}}, \bibinfo {author} {\bibfnamefont {Pina}\ \bibnamefont
  {Romaniello}}, \ and\ \bibinfo {author} {\bibfnamefont {Lucia}\ \bibnamefont
  {Reining}},\ }\bibfield  {title} {\enquote {\bibinfo {title} {Approximations
  for many-body green's functions: insights from the fundamental equations},}\
  }\href {\doibase 10.1088/1367-2630/14/1/013056} {\bibfield  {journal}
  {\bibinfo  {journal} {New Journal of Physics}\ }\textbf {\bibinfo {volume}
  {14}},\ \bibinfo {pages} {013056} (\bibinfo {year} {2012})}\BibitemShut
  {NoStop}%
\bibitem [{\citenamefont {Berger}\ \emph {et~al.}(2014)\citenamefont {Berger},
  \citenamefont {Romaniello}, \citenamefont {Tandetzky}, \citenamefont
  {Mendoza}, \citenamefont {Brouder},\ and\ \citenamefont
  {Reining}}]{Berger_2014}%
  \BibitemOpen
  \bibfield  {author} {\bibinfo {author} {\bibfnamefont {J~A}\ \bibnamefont
  {Berger}}, \bibinfo {author} {\bibfnamefont {Pina}\ \bibnamefont
  {Romaniello}}, \bibinfo {author} {\bibfnamefont {Falk}\ \bibnamefont
  {Tandetzky}}, \bibinfo {author} {\bibfnamefont {Bernardo~S}\ \bibnamefont
  {Mendoza}}, \bibinfo {author} {\bibfnamefont {Christian}\ \bibnamefont
  {Brouder}}, \ and\ \bibinfo {author} {\bibfnamefont {Lucia}\ \bibnamefont
  {Reining}},\ }\bibfield  {title} {\enquote {\bibinfo {title} {Solution to the
  many-body problem in one point},}\ }\href {\doibase
  10.1088/1367-2630/16/11/113025} {\bibfield  {journal} {\bibinfo  {journal}
  {New Journal of Physics}\ }\textbf {\bibinfo {volume} {16}},\ \bibinfo
  {pages} {113025} (\bibinfo {year} {2014})}\BibitemShut {NoStop}%
\bibitem [{\citenamefont {Di~Sabatino}\ \emph {et~al.}(2019)\citenamefont
  {Di~Sabatino}, \citenamefont {Berger},\ and\ \citenamefont
  {Romaniello}}]{stefano_JCTC}%
  \BibitemOpen
  \bibfield  {author} {\bibinfo {author} {\bibfnamefont {Stefano}\ \bibnamefont
  {Di~Sabatino}}, \bibinfo {author} {\bibfnamefont {J.~A.}\ \bibnamefont
  {Berger}}, \ and\ \bibinfo {author} {\bibfnamefont {Pina}\ \bibnamefont
  {Romaniello}},\ }\bibfield  {title} {\enquote {\bibinfo {title} {Many-body
  effective energy theory: photoemission at strong correlation},}\ }\href
  {\doibase 10.1021/acs.jctc.9b00427} {\bibfield  {journal} {\bibinfo
  {journal} {J. Chem. Theory Comput.}\ }\textbf {\bibinfo {volume} {15}},\
  \bibinfo {pages} {5080} (\bibinfo {year} {2019})}\BibitemShut {NoStop}%
\bibitem [{\citenamefont {Di~Sabatino}()}]{stefano_private}%
  \BibitemOpen
  \bibfield  {author} {\bibinfo {author} {\bibfnamefont {S.}~\bibnamefont
  {Di~Sabatino}},\ }\href@noop {} {}\bibinfo {howpublished} {private
  communication}\BibitemShut {NoStop}%
\bibitem [{Note1()}]{Note1}%
  \BibitemOpen
  \bibinfo {note} {Note that we have chosen this symmetric form for the matrix
  $\protect \mathbf {S}^v$, instead of ${S}^R_{ij}=n_i\delta _{ij}$, to keep
  contact with the original derivation by Morrell \protect \textit {et al.}
  \cite {morrell_JCP1975}. These two choices for $\protect \mathbf {S}^R$ yield
  the same eigenvalues for the matrix $\protect \mathbf {\Lambda
  }^R$.}\BibitemShut {Stop}%
\bibitem [{\citenamefont {Balzer}\ \emph {et~al.}(2011)\citenamefont {Balzer},
  \citenamefont {Gdaniec},\ and\ \citenamefont {Potthoff}}]{Balzer_2011}%
  \BibitemOpen
  \bibfield  {author} {\bibinfo {author} {\bibfnamefont {Matthias}\
  \bibnamefont {Balzer}}, \bibinfo {author} {\bibfnamefont {Nadine}\
  \bibnamefont {Gdaniec}}, \ and\ \bibinfo {author} {\bibfnamefont {Michael}\
  \bibnamefont {Potthoff}},\ }\bibfield  {title} {\enquote {\bibinfo {title}
  {Krylov-space approach to the equilibrium and nonequilibrium single-particle
  green's function},}\ }\href {\doibase 10.1088/0953-8984/24/3/035603}
  {\bibfield  {journal} {\bibinfo  {journal} {Journal of Physics: Condensed
  Matter}\ }\textbf {\bibinfo {volume} {24}},\ \bibinfo {pages} {035603}
  (\bibinfo {year} {2011})}\BibitemShut {NoStop}%
\bibitem [{\citenamefont {Gilbert}(1975{\natexlab{b}})}]{Gilbert75}%
  \BibitemOpen
  \bibfield  {author} {\bibinfo {author} {\bibfnamefont {T.~L.}\ \bibnamefont
  {Gilbert}},\ }\bibfield  {title} {\enquote {\bibinfo {title}
  {Hohenberg-{Kohn} theorem for nonlocal external potentials},}\ }\href
  {\doibase 10.1103/PhysRevB.12.2111} {\bibfield  {journal} {\bibinfo
  {journal} {Phys. Rev. B}\ }\textbf {\bibinfo {volume} {12}},\ \bibinfo
  {pages} {2111--2120} (\bibinfo {year} {1975}{\natexlab{b}})}\BibitemShut
  {NoStop}%
\bibitem [{\citenamefont {Sharma}\ \emph {et~al.}(2013)\citenamefont {Sharma},
  \citenamefont {Dewhurst}, \citenamefont {Shallcross},\ and\ \citenamefont
  {Gross}}]{Sharma13}%
  \BibitemOpen
  \bibfield  {author} {\bibinfo {author} {\bibfnamefont {S.}~\bibnamefont
  {Sharma}}, \bibinfo {author} {\bibfnamefont {J.~K.}\ \bibnamefont
  {Dewhurst}}, \bibinfo {author} {\bibfnamefont {S.}~\bibnamefont
  {Shallcross}}, \ and\ \bibinfo {author} {\bibfnamefont {E.~K.~U.}\
  \bibnamefont {Gross}},\ }\bibfield  {title} {\enquote {\bibinfo {title}
  {Spectral density and metal-insulator phase transition in mott insulators
  within reduced density matrix functional theory},}\ }\href {\doibase
  10.1103/PhysRevLett.110.116403} {\bibfield  {journal} {\bibinfo  {journal}
  {Phys. Rev. Lett.}\ }\textbf {\bibinfo {volume} {110}},\ \bibinfo {pages}
  {116403} (\bibinfo {year} {2013})}\BibitemShut {NoStop}%
\bibitem [{Elk()}]{elk}%
  \BibitemOpen
  Elk,\ \href@noop {} {} (\bibinfo {year} {Elk, http://elk.sourceforge.net,
  2004})\BibitemShut {NoStop}%
\bibitem [{\citenamefont {Sharma}\ \emph {et~al.}(2008)\citenamefont {Sharma},
  \citenamefont {Dewhurst}, \citenamefont {Lathiotakis},\ and\ \citenamefont
  {Gross}}]{sharma_PRB08}%
  \BibitemOpen
  \bibfield  {author} {\bibinfo {author} {\bibfnamefont {S.}~\bibnamefont
  {Sharma}}, \bibinfo {author} {\bibfnamefont {J.~K.}\ \bibnamefont
  {Dewhurst}}, \bibinfo {author} {\bibfnamefont {N.~N.}\ \bibnamefont
  {Lathiotakis}}, \ and\ \bibinfo {author} {\bibfnamefont {E.~K.~U.}\
  \bibnamefont {Gross}},\ }\bibfield  {title} {\enquote {\bibinfo {title}
  {Reduced density matrix functional for many-electron systems},}\ }\href
  {\doibase 10.1103/PhysRevB.78.201103} {\bibfield  {journal} {\bibinfo
  {journal} {Phys. Rev. B}\ }\textbf {\bibinfo {volume} {78}},\ \bibinfo
  {pages} {201103} (\bibinfo {year} {2008})}\BibitemShut {NoStop}%
\bibitem [{\citenamefont {Sze}(1969)}]{Semiconductor_devices}%
  \BibitemOpen
  \bibfield  {author} {\bibinfo {author} {\bibfnamefont {S.~M.}\ \bibnamefont
  {Sze}},\ }\href@noop {} {\emph {\bibinfo {title} {The Physics of
  Semiconductor Devices}}}\ (\bibinfo  {publisher} {Wiley},\ \bibinfo {address}
  {New York},\ \bibinfo {year} {1969})\BibitemShut {NoStop}%
\end{thebibliography}%

\end{document}